\documentclass[showpacs,preprintnumbers,amsmath,amssymb]{revtex4}
\usepackage{graphicx}

\begin{document}

\title {Angular velocity variations and stability of spatially explicit prey-predator systems}
\author{Refael Abta  and Nadav M. Shnerb}

\affiliation{   Department of Physics, Bar-Ilan University,
Ramat-Gan 52900 Israel}

\begin{abstract}
The linear instability of Lotka-Volterra orbits in the homogenous
manifold of a  two-patch system is analyzed. The origin of these
orbits instability in the absence of prey migration is revealed to
be the dependence of the angular velocity on the azimuthal angle; in
particular, the system desynchronizes at the exit from the slow part
of the trajectory. Using this insight, an analogous model of  a two
coupled oscillator is presented and shown to yield the same type of
linear instability. This  unables one to incorporate the linear
instability within a recently presented general framework that
allows for comparison of all known stabilization mechanisms and for
simple classification of observed  oscillations.
\end{abstract}

\pacs{05.45.-a, 05.45.Xt, 87.23.Cc, 02.50.Ey}

\maketitle

Population oscillations in prey predator systems, as predicted by
the Lotka-Volterra equations \cite{lotka}, are known to be unstable
with respect to additive and multiplicative  noise. This instability
must lead to the extinction of one of the interacting species, a
fact that has been confirmed in various experiments for well-mixed
populations \cite{gause}. The persistence of natural prey-predator
and host-parasitoid systems, thus, is commonly attributed to their
spatial structure, such that migration between desynchronized
patches yields an inward flow toward the coexistence fixed point and
is responsible for the sustainability of the oscillations
\cite{nic33}. In fact, spatially extended systems tend to support
finite amplitude oscillation \cite{kerr}. The stabilization of such
oscillations is considered to be a major factor affecting species
conservation and ecological balance \cite{levin}.

The main challenge, thus, is to understand the conditions for the
appearance of desynchronization in diffusively coupled patches,
since diffusion tends to synchronize these patches so after a while
the whole system flows to the  well mixed, unstable limit
\cite{hoopes}. One of the solutions to that problem was presented by
Jansen \cite{jansen,roos,zig}. It turns out that the trajectories
far from the fixed point become unstable if the inter-patch
migration rate of the predator is much larger than that of the prey.
Jansen used Floquet analysis to show that instability. In this
paper, we try to elucidate the underlying mechanism that yields
Jansen's instability, to generalize it in the framework of the
recently presented coupled oscillator model, and to discuss the
conditions under which one may observe the stabilizing effect of
Jansen's mechanism, like oscillation amplitude that grows under
noise until it reaches the  first unstable orbit.

The Lotka-Volterra predator-prey system is a paradigmatic model for
oscillations in population dynamics \cite{lotka}. It describes the
time evolution of  two interacting populations: a prey ($b$)
population that grows with a constant birth rate $\sigma$ in the
absence of a predator (the energy resources consumed by the prey are
assumed to be inexhaustible), while the predator population ($a$)
decays (with death rate $\mu$),  in the absence of a prey. Upon
encounter, the predator may consume the prey with  a certain
probability. Following a consumption event,  the predator population
grows and the prey population decreases. For a well-mixed
population, the corresponding partial differential equations are,
\begin{eqnarray}\label{basic}
\frac{\partial a}{\partial t} &=& - \mu a + \lambda_1 a b \\
\nonumber \frac{\partial b}{\partial t} &=& \sigma b - \lambda_2 a
b,
\end{eqnarray}
where $\lambda_1$ and $\lambda_2$ are the relative increase
(decrease) of the predator (prey) populations due to the interaction
between species, correspondingly.

The system admits two unstable fixed points: the absorbing state
$a=b=0$ and the state $a=0, \ \ b = \infty$. There is one marginally
stable fixed point at $\bar{a}  = \sigma/\lambda_2, \ \ \bar{b} =
\mu/\lambda_1$. Local stability analysis yields the eigenvalues $\pm
 i \sqrt{\mu \sigma}$ for the stability matrix. Moreover, even
 beyond the linear regime there is neither convergence nor
 repulsion. Using logarithmic variables $z = ln(a), \  q = ln(b)$
 eqs. (\ref{basic}) take the canonical form $\dot{z} = \partial
 H/\partial q, \ \ \dot{q} = -\partial
 H/\partial z$, where the conserved quantity $H$ (in the $ab$
 representation) is,
 \begin{equation}\label{H}
 H = \lambda_1 b + \lambda_2 a - \mu \ ln(a) - \sigma \  ln(b).
 \end{equation}
The phase space, thus, is segregated into a collection of nested
one-dimensional trajectories, where each one is characterized by a
different value of  $H$, as illustrated in Figure 3.  Given a line
connecting the fixed point to one of the "walls" (e.g., the dashed
line in the phase space portrait, Figure 3), $H$ is a monotonic
function on that line, taking its minimum $H_{min}$ at the
marginally stable fixed point (center) and diverging on the wall.
Without loss of generality, we employ hereon the symmetric
parameters $\mu = \sigma = \lambda_1 = \lambda_2 =1$. The
corresponding phase space, along with the dependence  of $H$ on  the
distance from the center and a plot of the oscillation period vs.
$H$, are represented in Figure 3).

\begin{figure}
  \includegraphics[width=8cm]{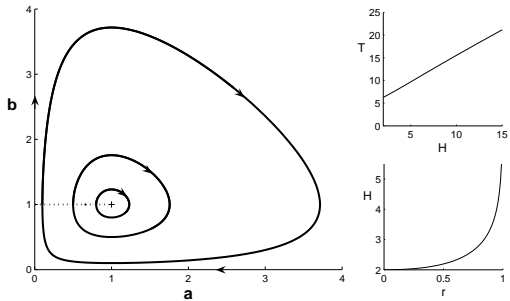}
  \caption{The Lotka-Volterra phase space (left panel) admits a
  marginally stable fixed point surrounded by close trajectories (three of these
  are
  plotted). Each trajectory corresponds to single $H$ defined in Eq.
  (\ref{H}), where $H$ increases monotonically  along the (dashed) line
  connecting the center with the $a=0$ wall, as shown in the lower
  right panel. In the upper right  panel, the period of a cycle $T$ is
  plotted against $H$, and is shown to increase almost linearly from
  its initial value $T = 2 \pi / \sqrt{\mu \sigma}$ close to the
  center.}
  \label{fig3}
\end{figure}

Given the integrability of that system, the effect of noise is quite
trivial: if $a$ and $b$ randomly fluctuate in time (e.g., by adding
or subtracting small amounts of population during each time step),
the system wanders between trajectories, thus performing some sort
of random walk in  $H$  with "repelling boundary conditions" at
$H_{min}$ and "absorbing boundary conditions" on the walls (as
negative densities are meaningless, the "death" of the system is
declared when the trajectory hits the zero population state for one
of the species). This result was emphasized by  Gillespie
\cite{Gillespie} for the important case where intrinsic stochastic
fluctuations are induced by the discrete character of the reactants.
In that case, the noise is multiplicative (proportional to the
number of particles), and the system flows away from the center and
eventually hits one of the absorbing states at $0,0$ or $0,\infty$.
The corresponding situation for a single patch Lotka-Volterra system
with additive noise is demonstrated in Figure 4, where the survival
probability $Q(t)$ (the probability that a trajectory does not hit
the absorbing walls within  time $t$) is shown for different noise
amplitudes.

\begin{figure}
  \includegraphics[width=7cm]{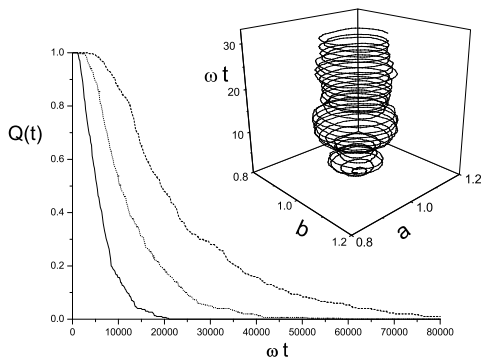}
 \caption{The survival probability $Q(t)$ is plotted versus time for
a single-patch, noisy LV
  system. Eqs. (\ref{basic}) (with the symmetric parameters) were integrated numerically (Euler integration
  with time step $0.001$), where the initial conditions are at the fixed point $a=b=1$. At each time step, a small random number
  $\eta(t) \Delta t$ was added to each population density, where $\eta(t) \in [-\Delta,\Delta]$. A typical phase space trajectory, for
  $\Delta = 0.5$, is shown in the inset. The system "dies" when the trajectory hits the walls $a=0$ or $b=0$. Using 300 different noise histories, the
  survival probability is shown here for $\Delta = 0.5$ (full line), $\Delta = 0.3$ (dotted line) and $\Delta = 0.25$ (dashed line).
  Clearly, the survival probability decays exponentially at long
  times, $Q(t) \sim exp(-t/\tau)$. As expected for a random walk
  with absorbing boundary conditions, $1/ \tau$ scales with
  $\Delta^2$.}
  \label{fig2}
\end{figure}

The Lotka-Volterra system on spatial domains has been investigated,
usually in a form of diffusively coupled patches, during the last
decades.  Any patch is assumed to be well mixed, and the flow of the
reactants from one patch to its neighbors is governed by the density
gradient. Clearly, any system of that type, independent of its
spatial topology (either regular lattice of some dimensionality or
some sort of network without isolated nodes) admits an infinite
number of solutions that correspond to \emph{synchronous}
oscillations of the whole system along one of the $H$ trajectories,
where the diffusion has no role as there are no population
gradients. The simplest example is  the two-patch system, described
by:

\begin{eqnarray}\label{two}
\frac{\partial a_1}{\partial t} &=& - \mu a_1 + \lambda_1  a_1 b_1 +
D_a (a_2-a_1) \\  \nonumber \frac{\partial a_2}{\partial t} &=& -
\mu a_2 + \lambda_1   a_2 b_2 + D_a (a_1-a_2)\\  \nonumber
\frac{\partial b_1}{\partial t} &=& \sigma b_1 - \lambda_2 a_1  b_1
+ D_b (b_2 - b_1) \\ \nonumber \frac{\partial b_2}{\partial t} &=&
\sigma b_2 - \lambda_2  a_2 b_2 + D_b (b_1 - b_2).
\end{eqnarray}

Here the invariant manifold is the two dimensional subspace
$a_1=a_2, \ \ b_1 = b_2$. The diffusion, of course, suppresses
fluctuations and stabilizes the invariant manifold; one may expect,
thus, that the single-patch dynamics also capture the main features
of the extended system, and that the system behaves like a random
walker in the invariant manifold (with a rescaled noise) and hits
the absorbing walls after some characteristic  time $\tau$, where
$\tau$ scales linearly with the noise strength $\Delta^2$.

As a first hint for a stabilizing mechanism, let us consider the
total $H$,
\begin{equation}\label{Htotal}\
H_{T} \equiv H_1 + H_2  = a_1 + a_2 + b_1 + b_2 - ln(a_1 a_2 b_1
b_2).
\end{equation}
With the deterministic dynamics (\ref{two}), $H_{T}$ is a
\emph{monotonously decreasing} quantity in the non-negative
population regime:
\begin{equation} \label{Hdynamics}
\frac{d H_T }{dt} = - D_a \left( \frac{(a_1 - a_2)^2}{a_1 a_2}
\right) - D_b \left( \frac{(b_1 - b_2)^2}{b_1 b_2} \right) < 0.
\end{equation}
Accordingly, if an orbit on the invariant manifold becomes unstable,
the flow will be inward and the population oscillations stabilizes.

\begin{figure}
  \includegraphics [width=8cm] {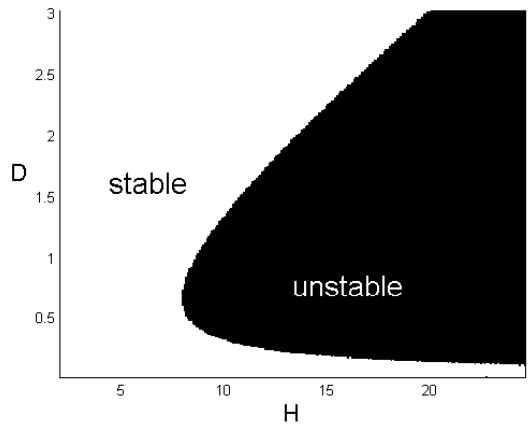}\\
  \caption{Stability diagram for phase space orbits (ordered by their conserved quantity $H$)
  for different values of predator diffusion $D_a$, where Db=0.}
  \label{fig1}
\end{figure}

While if $D_a = D_b$ the stability properties of an orbit on the
invariant manifold are identical to the stability properties of the
corresponding single-patch orbit \cite{Abarbanel}, if the diffusion
of both species is different, there is a possibility for unstable
orbits on the homogenous plane. This option was materialized by
Jansen \cite{jansen}, who considered the set of Equations \ref{two}
in the limit $D_b = 0$, so that only the predator undergos
diffusion. With the transformation:
\begin{eqnarray}\label{transform}
A=\frac{a_1 + a_2}{2} \qquad B=\frac{b_1 + b_2}{2}\\\nonumber
\delta=\frac{a_1 - a_2}{2} \qquad \theta=\frac{b_1 - b_2}{2},
\end{eqnarray}
one recognizes the homogenous $AB$ manifold and that  the $\delta$
$\theta$ coordinates measure the deviation from that manifold (the
heterogeneity of the population). In these coordinates the system
satisfies,
\begin{eqnarray}\label{lv2trans}
\frac{\partial A}{\partial t} &=& - \mu A + \lambda_1  A B + \lambda_1  \delta \theta\\
\nonumber \frac{\partial B}{\partial t} &=& \sigma B - \lambda_2
A  B + \lambda_2  \delta \theta \\
\nonumber \frac{\partial \delta}{\partial t} &=& - \mu \delta +
\lambda_1 A \theta + \lambda_1 B \delta -2 D_a \delta \\ \nonumber
\frac{\partial \theta}{\partial t} &=& \sigma B - \lambda_2  A
\theta - \lambda_2  B \delta -2 D_b \theta.
\end{eqnarray}
Linearizing around the homogenous manifold, The $AB$ dynamic is
equivalent to that of a  single patch,
\begin{eqnarray}\label{lv2translin}
\dot {A} &=& - \mu A + \lambda_1  A B\\
\nonumber \dot{B} &=& \sigma B - \lambda_2 A  B  \\ \nonumber
\end{eqnarray}
and the $\delta-\theta$ linearized dynamic is
\begin{eqnarray}\label{matrix}
 \frac{\partial}{\partial t} \left(
  \begin{array}{cc}
    \delta \\ \theta \\
  \end{array}
\right) =\left(
  \begin{array}{cc}
    -\mu+\lambda_1 B -2D_a & \lambda_1 A \\
    -\lambda_2 B & \sigma - \lambda_2 A -2D_b \\
  \end{array}
\right) \left(
  \begin{array}{cc}
    \delta \\ \theta \\
  \end{array}
\right).
\end{eqnarray}
One may thus calculate the eigenvalues of the Floquet operator for
one period along an orbit of the homogenous manifold
(\ref{lv2translin}). The resulting stability diagram, first obtained
by \cite{jansen}, is shown in Figure \ref{fig1}.

\begin{figure}
  \includegraphics [width=8cm] {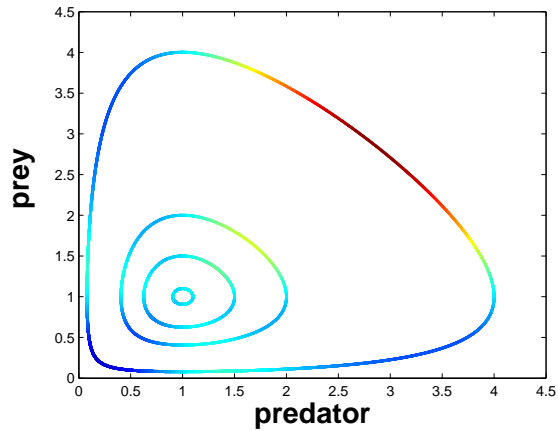}\\
  \caption{The angular velocity along some orbits of the  Lotka-Volterra
  dynamic. Fast regions marked in red, slow regions are blue.
   Clearly, the dynamics is slowest when the populations of both species are diluted, and fastest along the
   dense region in the upper-right "shoulder." Note that the velocity gradient along an orbit increases with $H$. }
  \label{ang}
\end{figure}

Our first mission is to intuitively explain  Jansen's results.
First, we notice that the angular velocity along a single
Lotka-Volterra orbit is not fixed. Figure \ref{ang} emphasizes the
angular velocity gradient along an orbit. While the inner
trajectories (close to the fixed point) are almost harmonic with
constant angular velocity, the eccentric large $H$ orbits admit
large variations. In particular, the motion in the dilute population
region [close to the unstable empty fixed point ($0,0$)] is very
slow, while in the dense population region the angular velocity is
large.

\begin{figure}
  \includegraphics [width=8cm] {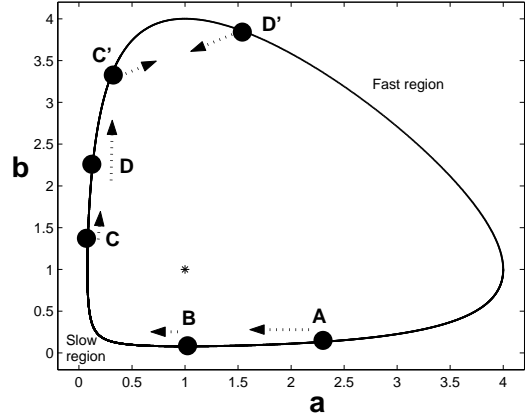}\\
  \caption{An orbit of the LV dynamics and its fast and slow regions. As explained in the text, with no prey migration  the
  two patches desynchronize in the $CD$ region, thus predator diffusion causes a flow toward the fixed point and stabilizes the oscillations.   }
  \label{z1}
\end{figure}

Following the caricature of an orbit in  Fig.  \ref{z1}, we can
explain the source of the instability. For a two-patch system, if
one patch is at point A along the orbit and the other patch at B,
since the A patch is moving faster along the line it will get closer
and closer to B during their flow toward the slow region. The
diffusion of the prey plays no role along this branch, since the
prey density is almost equal, while the predator diffusion may only
strengthen that effect. Thus, the two patches must (almost)
synchronize along this branch.

\begin{figure}
  \includegraphics [width=8cm] {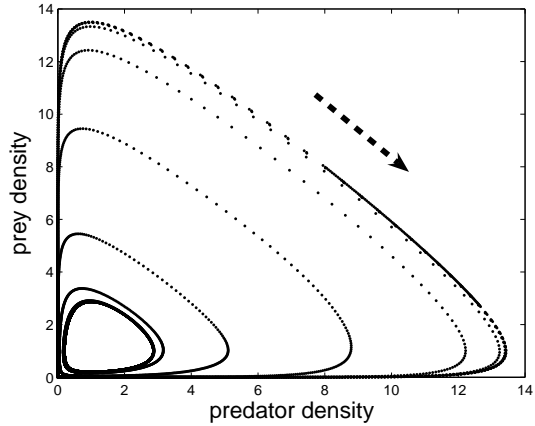}\\
  \caption{Phase portrait of the inward  flow in the homogenous manifold (average prey density vs. average predator density)
  for two-patch LV system with no prey diffusion and $D_{predators} = 1$. Clearly, the inward flow happens in the $C' - D'$ region of
  Figure \ref{z1}, where the desynchronization along the $CD$ branch interferes with the predator diffusion. There is almost no inward motion along the rest of the
  orbit.    }
  \label{zzj}
\end{figure}

The situation is completely different in the exit from the slow
region. The patch at D moves much faster than that at C, so they
will \emph{desynchronize}. As the predator density along this branch
is almost constant, the only factor that may avoid desynchronization
is the prey migration. In the absence of prey migration, the two
patches reach the points C' and D', where the predator migration
produces an inward flow. Figure \ref{zzj} is now well understood:
the inward flow happens when the desynchronization interferes with
the predator diffusion, as explained.

Let us consider, now, Jansen's instability in the framework of the
coupled nonlinear oscillator toy model, recently presented \cite{us}
as a generic tool for the investigation of oscillation stability in
diffusively coupled metapopulations. With the intuition gathered
from the above example, we want to consider diffusively coupled
orbits where the angular velocity depends on the radial angle and
the diffusing species density is changing along the slowing branch.
The following equations,
\begin{eqnarray}\label{co2}
 \frac{\partial x_1}{\partial t} &=&  \omega (\theta_1) y_1 + D_x (x_2-x_1)\\
  \nonumber \frac{\partial x_2}{\partial t} &=&  \omega (\theta_2) y_2 + D_x (x_1-x_2) \\
\nonumber \frac{\partial y_1}{\partial t} &=&  -\omega (\theta_1) x_1 + D_y(y_2-y_1) \\
 \nonumber \frac{\partial y_2}{\partial t} &=&  - \omega (\theta_2) x_2 + D_y (y_1
 -y_2),
 \end{eqnarray}
will satisfy these conditions for $D_x = D, \  D_y = 0$ and
\begin{eqnarray}\label{omegadef}
 \omega &=& \omega_0 + \omega_1 \cos(\theta-\frac{\pi}{4}).
  \end{eqnarray}

Using ($i \in 1,2$)
\begin{eqnarray}\label{tran} r_i^2 = x_i^2 + y_i^2 &\qquad&
  \theta_i  = arctan (\frac{y_i}{x_i}) \\
 \dot{r} = \frac {(x \dot{x} + y \dot{y})}{r} &\qquad&
 \dot{\theta} = \frac { (x \dot{y} - y \dot{x})}{r^2}, \nonumber
\end{eqnarray}
and
\begin{eqnarray}\label{transformation}
 r = r_2-r_1  &\qquad&
  R = r_2+r_1\\
 \phi = \theta_2-\theta_1 &\qquad&
  \Phi = \theta_2+\theta_1, \nonumber
\end{eqnarray}
one finds that the flow in the invariant manifold satisfies:
\begin{eqnarray}\label{proximity2}
  \dot R &=& 0 \\
  \nonumber \dot\Phi &=& \omega(\theta_1) + \omega(\theta_2) \\  \nonumber
  \end{eqnarray}
while the linearized equations for the desynchronization amplitude
$r$ and the desynchronization angle $\phi$ satisfy:

\begin{eqnarray}\label{mat}
\frac{\partial}{\partial t} \left(
  \begin{array}{cc}
    \phi \\ r \\
  \end{array}
\right)= \left(
  \begin{array}{cc}
    2\omega'(\Phi/2)-2D_x \cos^2(\Phi/2) & \frac{D_x \sin\Phi}{R} \\
    \frac{-D_x R\sin(\Phi)}{2} & -2D_x \sin^2(\Phi/2) \\
  \end{array}
\right)\left(
  \begin{array}{cc}
    \phi \\ r \\
  \end{array}
\right)
 \end{eqnarray}

Using the Floquet operator technique  to analyze the stability of an
 orbit by integrating (\ref{mat}) along a close trajectory of
 (\ref{proximity2}) one finds the stability map presented in Figure
 \ref{toy}, where the parameter $H$ of Figure \ref{fig1} is now
 replaced by $\omega_1$, which measures the "eccentricity" of the
 angular velocity along a circular path. Here, two unstable regions
 appear, for large and small $D_x$.

\begin{figure}[h]
\begin{center}
$\begin{array}{c@{\hspace{1in}}c} \multicolumn{1}{l}{\mbox{\bf (a)}}
&
    \multicolumn{1}{l}{\mbox{\bf (b)}} \\ [-0.53cm]
\includegraphics [width=8cm] {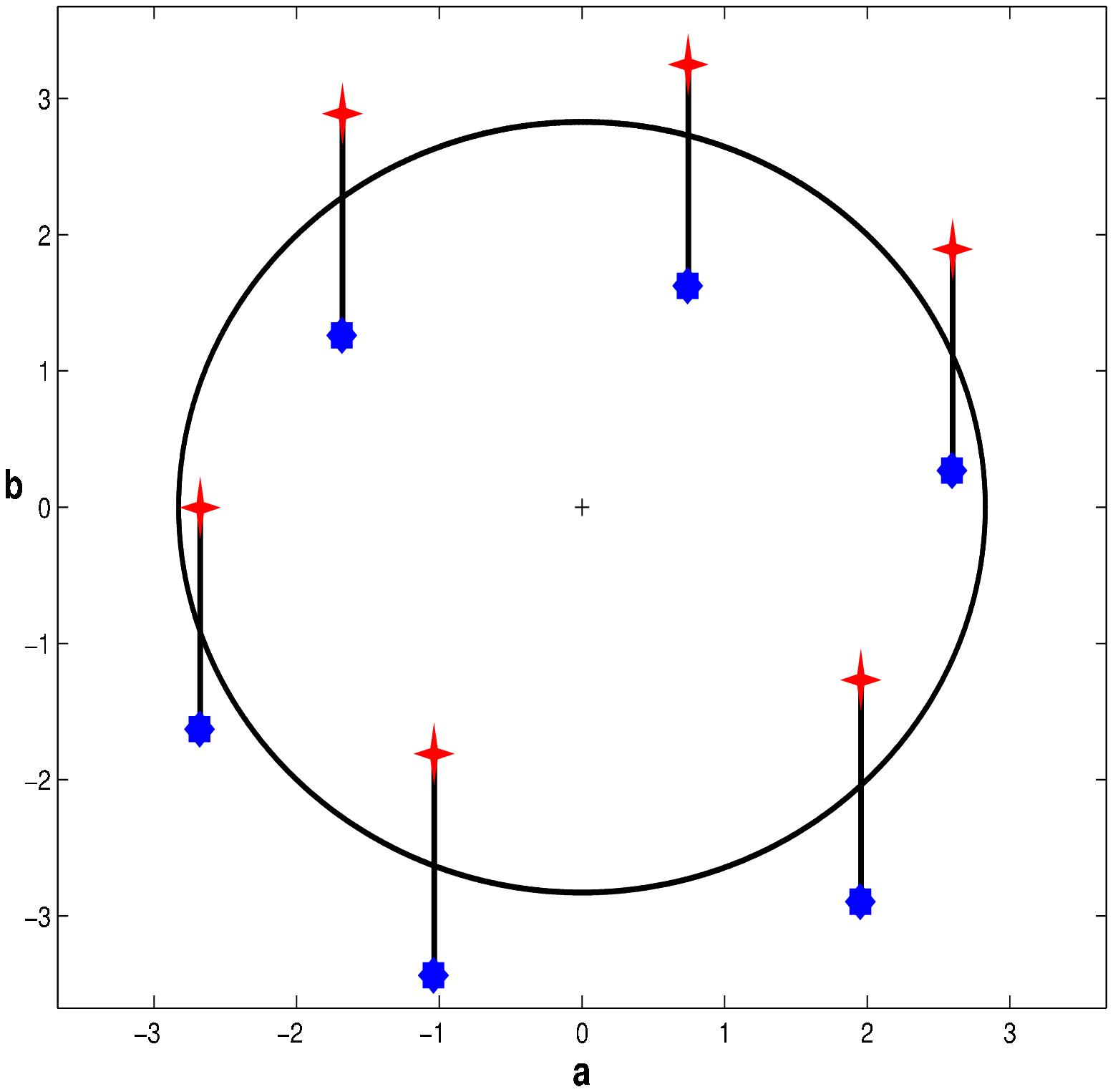} &
    \includegraphics [width=8cm] {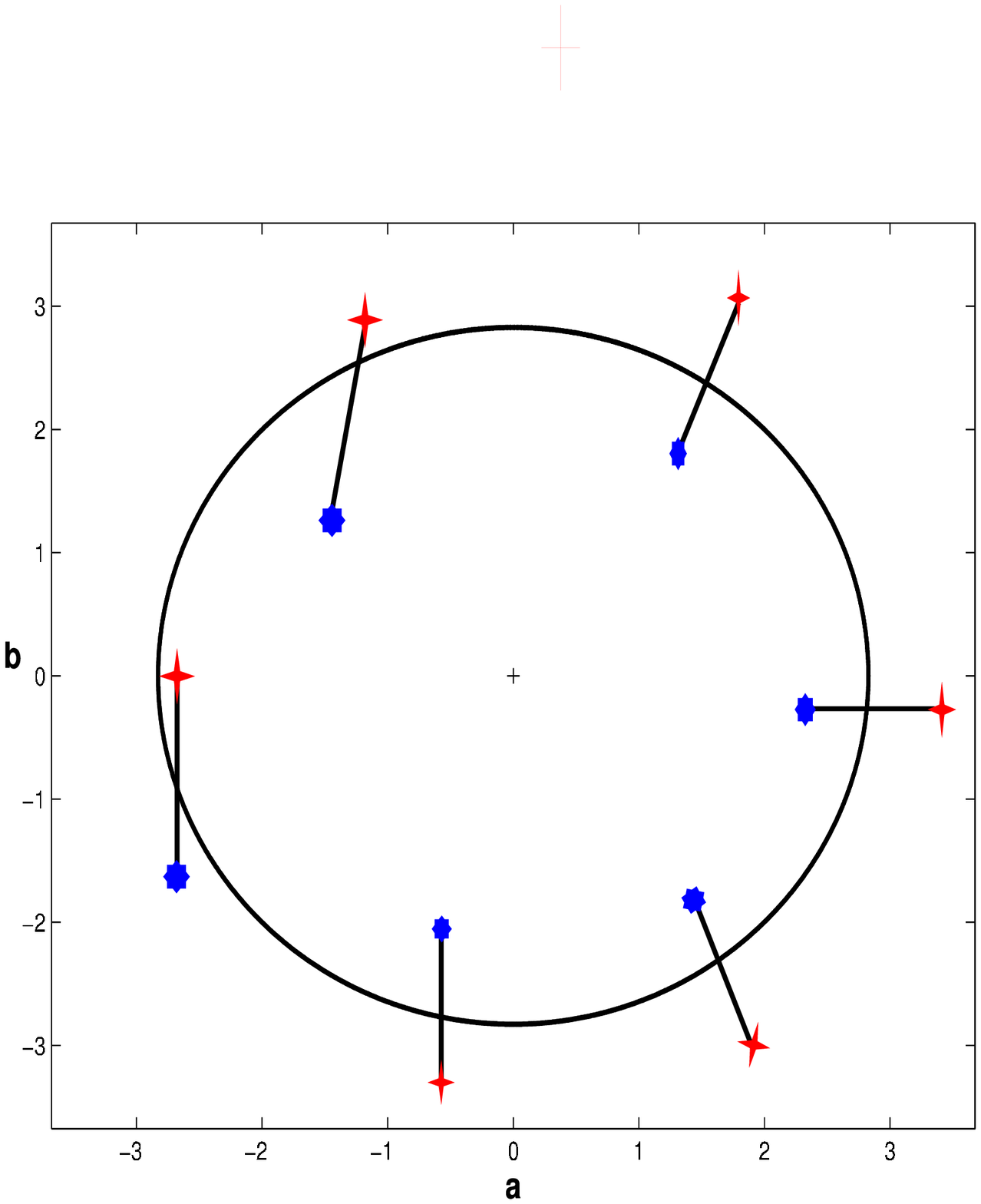}
\end{array}$
\end{center}
\caption{An illustration of the trajectories of two
diffusively-coupled patches, with slightly different
  initial conditions, projected on the invariant manifold.  In the strong coupling case, (left
  panel) the strong predator diffusion forces the two points to be
  on the same vertical line (same predator concentration) along the
  orbit, hence the phase of the Floquet eigenvalue inverted twice
  along the trajectory, yielding a positive eigenvalue. In the small
  diffusion limit, the patches posses equal predator density only in
  the slow portion of the orbit, when the intra-patch dynamic is
  slow with regard to the migration. This leads to trajectories like
  those illustrated in the right panel (points connected by "rod"
  stand for the population density in equal times), where only one
  sign change happens and the Floquet eigenvalue is positive.}
\label{rod}
\end{figure}

\begin{figure}
  \includegraphics [width=8cm] {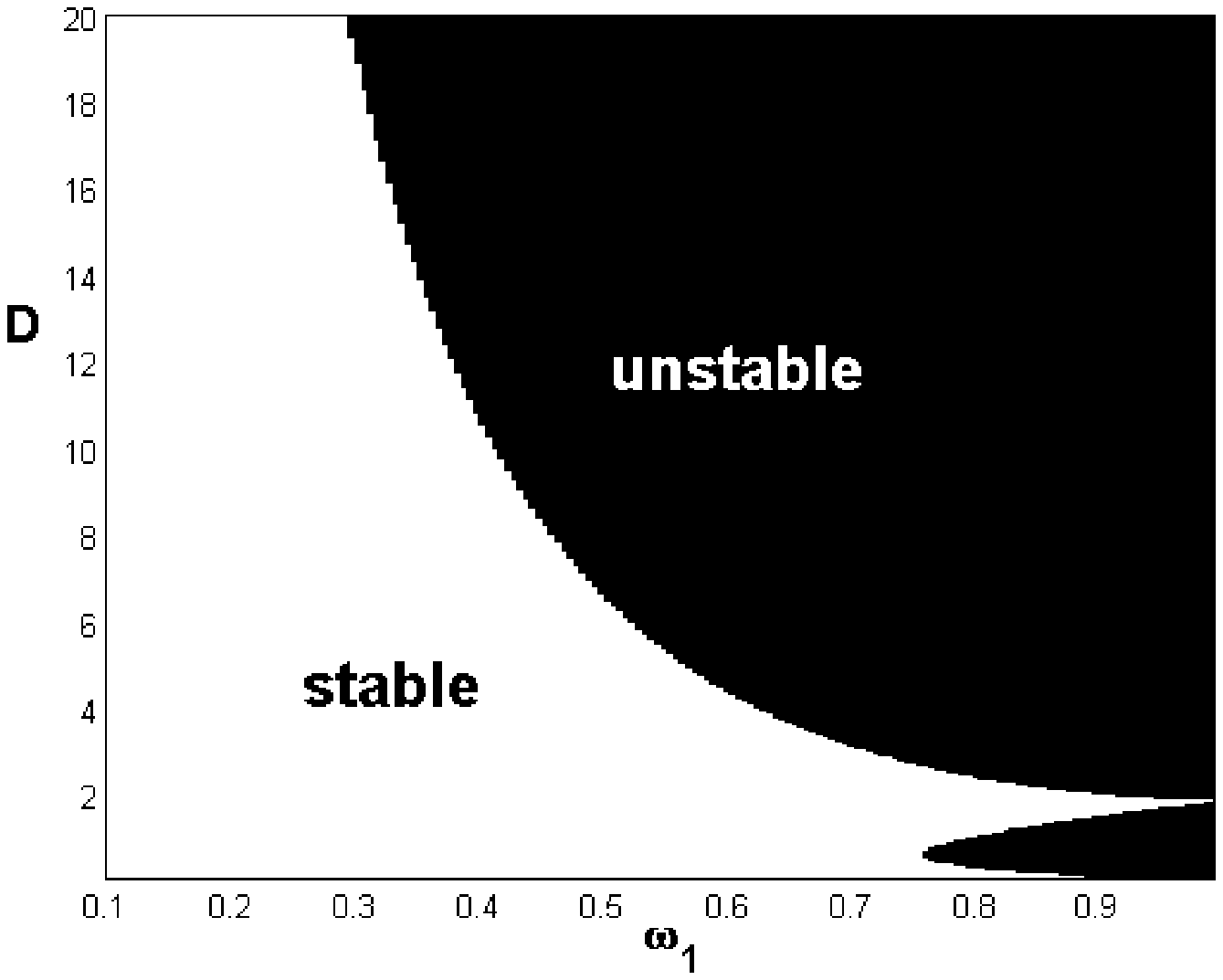}\\
  \caption{Stability diagram in the $\omega_1-D_x$ plane for the Floquet operator (same  as Figure \ref{fig1})
   for the coupled oscillator system described by Eqs. (\ref{co2})
   with $D_y = 0$. The two unstable regions correspond to different
   signs
   of the Floquet unstable eigenvalue, as explained in the text.}
  \label{toy}
\end{figure}

 It is interesting to point out that in the high $D_x$ region, the unstable  eigenvalue
 is positive, while in the small $D_x$ unstable region it takes
 negative values. The reason for that is the effect of predator
 migration. If the effect of migration is large, in comparison with
 the intra-patch dynamics, the two patches should admit (almost) the same
 predator density, and the corresponding points in the 2d phase
 portrait  should stay on the same
 vertical  line (same "x" coordinate). The trajectories of the two
 points representing the patches along an orbit are thus similar to
 the transport of a vertical rod along a circle, keeping the center
 of the rod on the circular trajectory: the two ends switch their
 role \emph{twice} (the inner becomes the outer and vice versa) so
 the rod returns to its original state after  a full cycle (See Figure \ref{rod}, left panel). In the
 low diffusion range, on the other hand, the points remain on a
 vertical line only close to the slow portion of the orbit, where
 they switch once, but in the fast region they support different
 predator populations, so the "rod" completes its cycle in opposite
 "phase"  (see Figure \ref{rod}, right panel).

We now turn to our last point, a comparison of this stabilizing
mechanism with the nonlinear, noise induced mechanism recently
presented by us \cite{us}. The stability mechanism of \cite{us}
involved with the \emph{amplitude dependence} of the angular
velocity (see the upper right panel of Figure \ref{fig3}), works as
well for system of equal prey and predator migration rates and is
not based on a linear instability of an orbit. One may ask, thus,
how to make a distinction between these two mechanisms in real
systems.

In order to make a distinction between amplitude-induced stability
\cite{us} and angular-induced stability \cite{jansen} one should
compare the corresponding radius of oscillations, where the dominant
mechanism corresponds to the smaller radius. The amplitude
synchronization prediction is that the oscillation radius scales
like $D/ ( \omega')^2$, where $\omega' \equiv \partial \omega /
\partial r$ is the frequency gradient along the  oscillations amplitude
(See Figure \ref{fig3}, upper right panel). This result should be
compared with the instability radius of \cite{jansen}, and for small
migration rates ($D \sim 0.01$)  it is smaller in few orders of
magnitude. It seems, thus, that the angular induced instability will
be relevant only for relatively large diffusivities, where the
effect of amplitude-induced instability is suppressed by patches
synchronization.

\begin{figure}[h]
\begin{center}
$\begin{array}{c@{\hspace{1in}}c} \multicolumn{1}{l}{\mbox{\bf (a)}}
&
    \multicolumn{1}{l}{\mbox{\bf (b)}} \\ [-0.53cm]
\includegraphics [width=8cm] {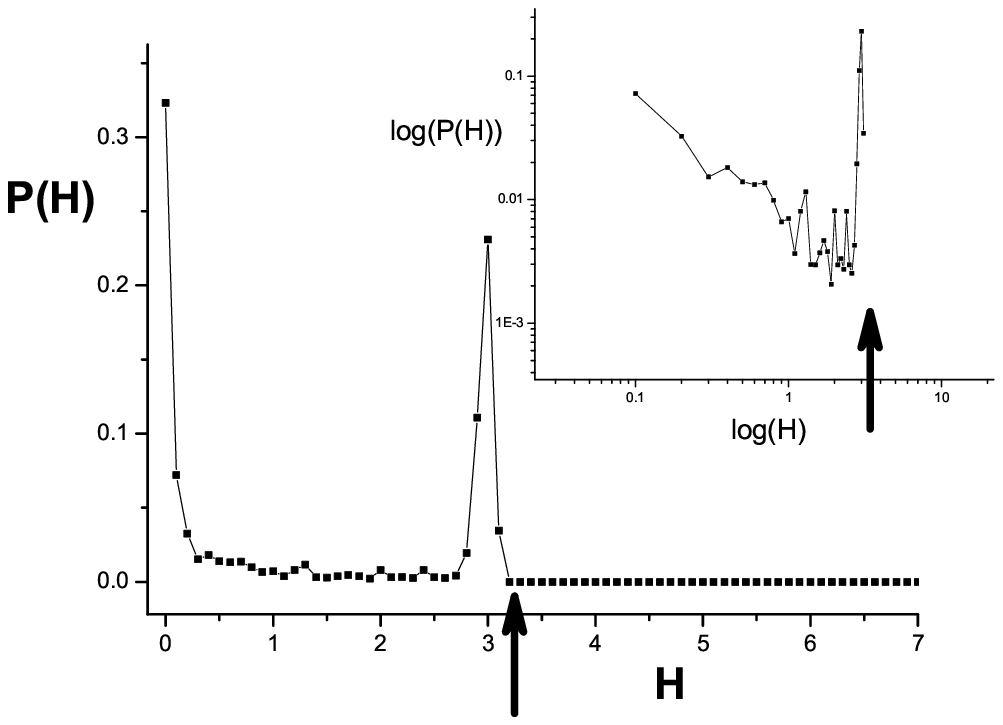} &
    \includegraphics [width=8cm] {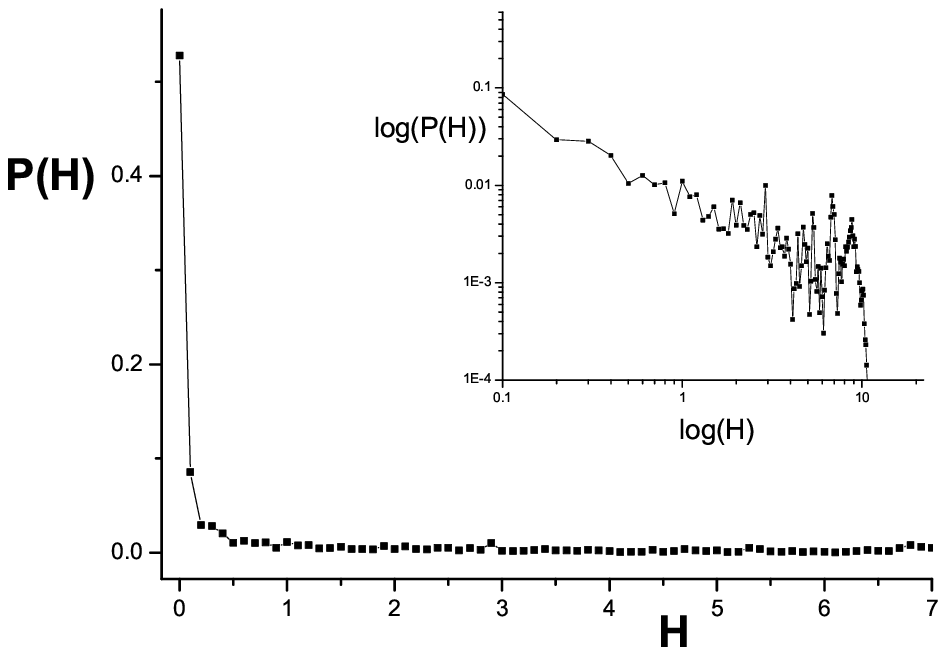}
\end{array}$
\end{center}
\caption{Histograms of the probability density as a function of $H$,
for a two-patch LV system with only
  prey diffusion  ($D_b = 1, \ D_a = 0$)(b) and only predator diffusion ($D_a = 1, \ D_b = 0$)(a). Both systems were subject to demographic stochasticity, modeled
  by a multiplicative noise proportional to the square root of the population density. In both cases, the probability density is concentrated
  around $H=0$; however, Jansen's instability manifests itself in the peak at the instability radius at the left panel, caused by the
  "reflection" from the unstable manifold (note the arrow that indicates the first unstable orbit). The log-log plots of that histogram
  (insets) show that the tail of the distribution is continuous at the right panel,
  but the probability to find the system with $H$ above the instability limit is  practically zero. The LV parameters are $\mu = \sigma =1$ and
  $\lambda = 10^{-5}$. }
\label{xyz}
\end{figure}

In order to observe the phase instability, we have simulated the
two-patch LV system, where the effect of  demographic stochasticity,
an intrinsic noise that should appear in any system independent of
environmental factors, was introduced via a noise term proportional
to the square root of the population size. The results are presented
in Figure \ref{xyz}, for the two opposite cases: no predator
migration (where one should expect angular instability), and no prey
diffusion, where no such instability is present. Both cases were
simulated for $\mu = \sigma = 1$, so the population (number of
individuals in each of the species)  at the coexistence fixed point
is $1/\lambda$. The diffusion $D=1$ corresponds to the smallest
amplitude of the last stable orbit  and is way too large to allow
amplitude desynchronization (see \cite{us}).

Fig. \ref{xyz} clearly shows the stabilizing effect of
angular-induced desynchronization.  For small populations at the
fixed point (large $\lambda$) the prey diffusion systems reach the
absorbing state, while only  predator diffusion stabilizes the inner
orbits. Large populations, though, may be stable in both regimes,
but the instability cuts the tail of the distribution, leaving only
a peak close to the "reflecting boundary."

To conclude, it has been  shown that systems where only the predator
admits the ability to migrate [a canonical examples include
herbivore - plant or parasite insect - plant systems, like in the
famous example of biological control of the Prickly Pear cactus by
the moth Cactoblastis cactorum in eastern Australia \cite{cac}] may
support sustained oscillations in noisy environments. This
phenomenon has been explained here, and its cause was traced to the
dependence of the angular velocity on the azimuthal angle along an
orbit lying in the homogenous manifold. This insight allows us to
incorporate that phenomenon into a generic framework of coupled
nonlinear oscillators and to compare that mechanism with other
stabilizing effects. In a separate publication \cite{us2}, we intend
to put forward a general classification scheme for stable population
oscillations, a scheme that may be used to typify the observed
desynchronization-induced stable manifold according to its
underlying mechanism.

\end{document}